# Toward a Curriculum Proposal for CSCW Education and Training in Latin America

Francisco J. Gutierrez
frgutier@dcc.uchile.cl
Department of Computer Science, University of Chile
Santiago, Chile

Yazmín Magallanes
yazmin.magallanesvz@udlap.mx
Interactive and Cooperative Technologies Lab, Universidad de las Américas
San Andrés Cholula, Puebla, Mexico

Laura S. Gaytán-Lugo
laura@ucol.mx
School of Mechanical and Electrical Engineering, Universidad de Colima
Coquimatlán, Colima, Mexico

Claudia López
claudia@inf.utfsm.cl
Universidad Técnica Federico Santa María
Valparaíso, Chile

Cleidson R. B. de Souza
cleidson.desouza@acm.org
Universidade Federal do Pará
Belém, PA, Brazil

## ABSTRACT

Computer-Supported Cooperative Work, or simply CSCW, is the research area that studies the design and use of socio-technical technology for supporting group work. CSCW has a long tradition in interdisciplinary work exploring technical, social, and theoretical challenges for the design of technologies to support cooperative and collaborative work and life activities. However, most of the research tradition, methods, and theories in the field follow a strong trend grounded in social and cultural aspects from North America and Western Europe. Therefore, it is inevitable that some of the underlying—and established—knowledge in the field will not be directly transferrable or applicable to other populations. This paper presents the results of an interview study conducted with Latin American faculty on the feasability, viability, and prospect of a curriculum proposal for CSCW Education in Latin America. To this end, we conducted nine interviews with faculty currently based in six countries of the region, aiming to understand how a CSCW course targeted to undergraduate and/or graduate students in Latin America might be deployed. Our findings suggest that there are specific traits that need to be addressed in such a course, such as: tailoring foundational CSCW concepts to the diversity of local cultures, motivating the involvement of students by tackling relevant problems to their local communities, and revitalizing CSCW research and practice in the continent.

## CCS CONCEPTS

• **Human-centered computing → Collaborative and social computing**; • **Social and professional topics → Computing education**; *Model curricula*.

## KEYWORDS

CSCW, Latin America, HCI Education and Training

## 1 INTRODUCTION

Computer-Supported Cooperative Work, or CSCW for short, is the interdisciplinary research area that studies design and use of technologies for cooperative and collaborative work as well as in every-day activities. These activities can be conducted in groups, organizations, communities, and even global networks. Given its long interdisciplinary tradition, it is no surprise that CSCW explores different facets of the dynamics around technology including technical, social, and theoretical aspects.

One of the broadly accepted findings of CSCW research is that collaborative technology use is often context-specific, i.e., depending on the context, people might use a system differently [10]. Furthermore, the norms for using a system are often actively negotiated and renegotiated among users [1]. However, since the term CSCW was coined by Irene Greif and Paul Cashman in 1984 [9], the field has been dominated by academic and industry research conducted in North America and Europe. Thus, most research—and literature—in the field has been produced in the northern hemisphere. Therefore, it is likely that key contextual aspects critical for understanding the use of technologies remains understudied due to the lack of global coverage of current CSCW research.

CSCW as a field has been studied in Latin America (LatAm), specially in countries like Brazil, Chile, and Mexico. However, their findings have not yet been broadly disseminated. Lopez et al. [13] have singled out a number of relevant aspects that can explain the relative invisibility of LatAm research, including the lack of a robust socio-technical research infrastructure, a trade-off between being locally impactful and globally relevant, and a language barrier. Nevertheless, exciting opportunities for LatAm researchers to contribute to the global research community have also been identified, in topics such as critical analysis on and beyond cultural differences and designing *"against the system"* to propose alternative configurations for current work and data practices.

With these opportunities in mind, we aim to extend state-of-the-art in LatAm-centric CSCW research by implementing various mechanisms to characterize CSCW research in the continent and strengthen this community. In that respect, this work addresses the need to learn about and envision CSCW in LatAm institutions. We report the results of an interview study conducted with faculty about a curriculum proposal for CSCW Education and Training in the region. We conducted nine interviews to faculty currently based in six countries in LatAm. The study aims to answer two research questions:

- (RQ1) *How does a CSCW undergraduate and/or graduate course in Latin America should look like?*
- (RQ2) *What would be the challenges and barriers to overcome for effectively implementing such a course?*

This study identifies a preliminary set of viewpoints and expectations on the viability and feasability of proposing an undergraduate/graduate CSCW course proposal, specifically tailored to LatAm.

In particular, the study findings highlight that this course needs to acknowledge the diversity of local cultures that make up our continent, as opposed to the mainstream visions of CSCW, which are largely represented in current literature and are based in the views and lifestyles of Anglo-American and Western European societies and cultures. In addition, an effective course proposal in LatAm should motivate the involvement of students by explicitly addressing relevant problems to their local communities.

The rest of this paper is structured as follows. Section 2 reviews and discusses related work. Section 3 presents the study methodology. Then, on Section 4 we present preliminary findings drawn from our empirical work. Finally, Section 5 concludes and provides perspectives on future work.

## 2 RELATED WORK

The Human-Computer Interaction (HCI) community has a long tradition of building curricular material to enhance and expand the field through education. Just before the 90's decade began, SIGCHI (i.e., ACM Special Interest Group on Computer-Human Interaction) started a series of panels which eventually became the *ACM Curricula for Human-Computer Interaction*, commonly known as the "Lime Green Report" [11]. This framework has been used by researchers, academics, and professionals to educate on subjects related to HCI. Even though that curriculum is still used, several authors argued on behalf of improving it to (1) update themes and (2) add a global perspective [5]. Therefore, initiatives such as SIGCHI Education Project [18] and the HCI Living Curriculum [6] emerged to fulfill these needs. In that respect, education professionals from various world regions have participated in the integration of different cultural perspectives through workshops held in events such as AfriCHI and CHI [12, 15, 19, 20].

Aligned with these efforts, HCI Collab is an Ibero-American network that aims to structure an HCI curricular proposal that works as a model for several institutions in the region. This innitiative seeks to counteract that HCI has not received full recognition within the countries in this region [7]. Other researchers in LatAm have proposed mechanisms for teaching and learning HCI through design and evaluation of computer games [17]. Likewise, the Mexican community has published two books to support HCI courses [4, 14].

More closely related to CSCW, two members of the Brazilian research community started in 2010 a collaborative project aiming to produce a book about CSCW to be used in undergraduate and graduate courses in Brazil. This book was eventually published in 2012 [16] with 26 chapters that where written by 49 Brazilian CSCW researchers. Chapters were organized in 5 different parts, namely: Fundamentals, Systems and Domains, Techniques, Development [of CSCW Systems], and Research [Methods]. Since this book was designed to be used in undergraduate and graduate classes in Brazil, it was written entirely in Portuguese. Interestingly, although some chapters use examples from the Brazilian context, none of them addresses something that is unique to Brazil.

Almost ten years after such a valuable contribution, we aim to contribute to CSCW Education and Training by exploring viewpoints and expectations of faculty working in LatAm institutions in regard to a CSCW curriculum proposal that focuses on the singularities of the LatAm context.

## 3 STUDY METHODOLOGY

In order to gather meaningful insights for better understanding the viewpoints, expectations, and main concerns around the development of a CSCW course curriculum tailored to LatAm, we conducted a semi-structured interview study with a sample of leading CSCW researchers currently based in the region.

### 3.1 Participants

Through snowball sampling, we recruited a total of nine active CSCW researchers currently based in LatAm and working in Academia. In particular, we reached participants from Mexico, Costa Rica, Colombia, Brazil, and Chile. In order to seed the sampling schema, we referred to the list of participants attending a specialized workshop held during the 2018 ACM CSCW Conference, focusing on the challenges and opportunities of conducting CSCW research in LatAm [8]. Table 1 describes study participants.

**Table 1: Study Participants**

| ID | Country | Main Field |
|---|---|---|
| P1 | Mexico | CSCW |
| P2 | Mexico | UbiComp |
| P3 | Mexico | HCI |
| P4 | Mexico | HCI – Engineering |
| P5 | Costa Rica | UbiComp – CSCW |
| P6 | Colombia | HCI – CSCW |
| P7 | Brazil | Software Engineering – CSCW |
| P8 | Chile | Software Engineering – CSCW |
| P9 | Chile | Information Retrieval – CSCW |

### 3.2 Data Collection

We conducted individual semi-structured interviews during April and May 2019. The interviews were conducted in-person, over the phone, or using videoconferencing services (e.g., Google Hangouts and Skype), depending on the availability and preference of each participant. Interviews lasted between 20 and 40 minutes and were conducted in Spanish and Portuguese depending on the participant.

We audio-recorded the interviews with the explicit consent from each participant, for later transcription and analysis. The interview script was validated in a pilot study, aiming to resolve wording problems and ambiguous statements. We collaboratively created the script in English, and then translated it to Spanish and Portuguese.

The main topics covered in the interview were: (1) *How does CSCW research conducted in LatAm compare to the American and European scenarios?*; (2) *How does the participant envision a LatAm- centric CSCW undergraduate and/or graduate course?*; and (3) *What challenges and barriers might emerge for putting into practice a LatAm-centric CSCW course curriculum?*.

### 3.3 Data Analysis

We generated our dataset by transcribing the collected audio data and later performing open, axial, and selective coding. To analyze our dataset, we followed the thematic analysis approach [2], which consists of generating initial codes from the data, searching for themes, contrasting the identified themes with the data, and iteratively refining the themes and narrative. Quotes, freely translated from Spanish and Portuguese by the authors, are provided to illustrate the main topics grouped under each theme.

## 4 FINDINGS

Based on a preliminary examination of the collected data, we structure our analysis in three main themes. The following subsections briefly present and discuss our interim results.

### 4.1 Current Status of CSCW Education and Training in Latin America

CSCW courses in LatAm mostly focus on students learning to do research in the area. Universities teach this elective course in the last quarter or semester of the undergraduate training program.

*"(My course) is about students writing a scientific article on a topic that interests them, and they afterwards conduct an applied project in that area."* —Participant P1.

Regarding the courses currently taught by the participants, they are usually delivered once a year in computer science and engineering departments. These courses are mainly targeted to undergraduate students enrolled in these departments' programs, although in some cases students from other areas such as electrical engineering and industrial engineering also participate. Although participation in some cases can reach up to 20 students, on average 17 students enroll in these classes. It is not common that these courses include graduate students.

Some of the interviewed researchers use practical examples taken from CSCW material developed by their collegues in other countries. However, given that most of the existing material in the field has been published with an underlying focus on Anglo-American and Western European societies and cultures, most LatAm instructors must adapt such a material to fit in our regional context.

*"I use references from the United States or Europe. Typically, I base my course on basic notions from seminal papers in the field, but I also have to translate (and adapt) the stated examples. I can't ask students to review a paper of an application in Europe and try to assimilate it because it does not usually apply in Latin America."*—Participant P5.

Overall, the study participants agreed that they teach the core concepts of CSCW and then focus on specific application domains in cooperative work, such as: learning, organizational collaboration, social computing, Web and mobile platforms, sentiment analysis, design and development of tools, indigenous communities, crowdsourcing, and techniques to organize masses to make decisions. Therefore, these focus areas can induce the development of materials, specifically tailored to LatAm, hence serving as a guide in the conception of a course.

### 4.2 Challenges and Barriers to Deploy a CSCW Course Tailored to Latin America

According to our informants, few universities offer CSCW courses in LatAm. During the interviews, the participants commented on the obstacles to delivering these courses at their universities. The obstacles mainly have to do with the structure of the universities, as well as a lack of CSCW communities in LatAm.

*4.2.1 Structure of Training Programs 2.* Some researchers praise the intrinsic cross-disciplinary nature of CSCW and HCI as a field, to enrich the interaction between students.

*"My elective course can be taken by students from, for example, social communication, graphic design, and anthropology, so we could have students with different visions."*—Participant P6.

However, despite successful cases, reality is that departments are usually geographically dispersed, which discourages collaboration and interaction among students coming from different training programs and disciplines. According to some of our participants, it is unlikely that LatAm universities collaborate with other universities, and even other departments within the same university, which is considered an enabling condition to teach engaging CSCW courses. The structure of the universities and the geographic infrastructure inhibit the interaction between the departments. For instance, participant P1 directly refers to Mexico's 1968 Student Movement [3]:

*"Government was very concerned that students could join together to do mischief, (...) so they created structures discouraging the interaction between departments"*—Participant P1.

To solve these drawbacks, our study participants look for ways to deliver CSCW courses like summer schools and hackathons. They also strategize through other projects to create links and organic collaborations with students from different departments and, in some cases, other universities within the same city. Another suggestion coming from participants is to develop mechanisms to teach CSCW courses for instructors of other areas, such as social science, or even software engineering.

*4.2.2 Revitalizing the Latin American CSCW Community.* Our study participants generally complained about a lack of communities that collaborate in the CSCW area in LatAm. Until a few years ago, there were more experts and multidisciplinary groups interested in CSCW in the continent. These groups lost strength and were dissolved. For instance, differences in time zones and language, specifically between Brazil and the rest of LatAm, pose a threat to collaboration among local CSCW communities. In that respect, the interviewed participants indicated that there is also a lack of material in Spanish and Portuguese specifically designed for LatAm.

*"We found that there was a lot of documentation in English, but we did not find resources in Spanish or Portuguese that our students could use."* —Participant P8.

Most of the CSCW material is proposed by other countries (e.g., United States and Great Britain), and this material is designed according to their needs. Therefore, our study participants suggest creating courses for LatAm contexts. For instance, in 2012 the Brazilian CSCW community published a textbook covering the main foundations in the field [16]. However, we see this situation as an exception, rather than the norm. For instance, this book does not address specific CSCW problems that are unique to Brazil, which makes using this material somewhat more difficult, when contextualizing and applying CSCW theory and design practices in LatAm-centric courses.

## 4.3 Implications to Course Design

Overall, participants agreed that designing a CSCW course, specifically tailored to LatAm, would be a valuable contribution to the community. In such a course, general—foundational—topics should be covered (e.g., awareness, participation, collaborative modeling), emphasizing the local and cultural components of the LatAm context.

In that respect, our study participants recommended covering more contemporary topics, such as: network and group formation strategies, gamification, science fiction, data science, design, requirements engineering, statistics for user studies, prototyping, crowdsourcing, ethnography, ethics, access to data, and teaching students to find gaps in research. Participants also commented on the importance of collaboration between students and communities.

## 4.4 Study Limitations

Although valuable, our current results are not exempt from limitations. Despite reaching saturation, we acknowledge that our sampling schema might not be complete at this point. In fact, our informants came only from Academia, which could potentially limit the variability of viewpoints and expectations when structuring a course outline aimed to train the next generation of CSCW researchers and practitioners. In that respect, our next steps in this research will aim to gather the views from industry and non-governmental organizations.

## 5 CONCLUSION AND FUTURE WORK

CSCW is the research area that studies the design and use of socio-technical technology for supporting group work, whereas such group work can be conducted by teams, organizations, communities, or networks. CSCW has a tradition in interdisciplinary work exploring technical, social, and theoretical aspects related to the design of technologies to support cooperative activities at work and/or home.

Most CSCW research is strongly grounded in social and cultural aspects from North America and Western Europe. Therefore, it is inevitable that some of the underlying—and established—knowledge in the field will not be directly transferrable or applicable to other populations. This paper presented an exploratory study to focus on this aspect, i.e., the feasability, viability, and prospect of a curriculum proposal for CSCW Education in LatAm. We interviewed nine faculty from six countries of the region, aiming to understand how a CSCW course targeted to undergraduate and/or graduate students in LatAm might be deployed. Our preliminary findings suggest that there are specific traits that need to be addressed in such a course, such as: (1) tailoring foundational CSCW concepts to the diversity of local cultures, (2) motivating the involvement of students by tackling relevant problems to their local communities, and (3) revitalizing CSCW research and practice in the continent.

As future work, we plan to conduct additional interviews with other faculty to understand whether there are new issues that were not reported by our study participants. Likewise, we will explicitly collect complementary viewpoints, such as those coming from industry partners and non-governmental organizations, as a way to have a richer view on the feasability and viability of proposing a LatAm-centric CSCW course proposal.

## REFERENCES

[1] Mark S. Ackerman. 2000. The intellectual challenge of CSCW: The gap between social requirements and technical feasibility. *Human-Computer Interaction* 15, 2 (2000), 179–203. https://doi.org/10.1207/S15327051HCI1523_5

[2] Virginia Braun and Victoria Clarke. 2006. Using thematic analysis in psychology. *Qualitative Research in Psychology* 3, 2 (2006), 77–101. https://doi.org/10.1191/ 1478088706qp063oa

[3] Claire Brewster. 2002. The Student Movement of 1968 and the Mexican Press: The Cases of "Excélsior" and "Siempre"! *Bulletin of Latin American Research* 21, 2 (2002), 171–190. http://www.jstor.org/stable/3339451

[4] Luis Castro and Marcela D. Rodríguez. 2019. *Interacción Humano-Computadora y Aplicaciones en México*. AMEXCOMP.

[5] Elizabeth F. Churchill, Anne Bowser, and Jennifer Preece. 2013. Teaching and Learning Human-computer Interaction: Past, Present, and Future. *Interactions* 20, 2 (March 2013), 44–53. https://doi.org/10.1145/2427076.2427086

[6] Elizabeth F. Churchill, Anne Bowser, and Jennifer Preece. 2016. The Future of HCI Education: A Flexible, Global, Living Curriculum. *Interactions* 2 (Feb. 2016), 70–73. https://doi.org/10.1145/2888574

[7] HCI Collab. 2019. Adecuación de la propuesta a la linea de acción del Área Temática correspondiente. Retrieved June 9, 2019 from http://hci-collab.com/

[8] Cleidson R. B. de Souza, Claudia López, Francisco J. Gutierrez, Laura S. Gaytán-Lugo, Marcos R. S. Borges, and Cecilia Aragon. 2018. Latin America As a Place for CSCW Research. In *Proceedings of the ACM Conference on Computer Supported Cooperative Work and Social Computing – Companion Volume (CSCW '18)*. ACM, New York, NY, USA, 401–407. https://doi.org/10.1145/3272973.3273015

[9] Jonathan Grudin and Steven Poltrock. 2012. Computer supported cooperative work. Retrieved June 7, 2019 from https://www.interaction-design.org/literature/book/the-encyclopedia-of-human-computer-interaction-2nd-ed/computer-supported-cooperative-work

[10] Jonathan Grudin and Steven E. Poltrock. 1997. Computer-supported cooperative work and groupware. *Advances in Computers* 45 (1997), 269–320. https://doi.org/ 10.1016/S0065-2458(08)60710-X

[11] Thomas T Hewett, Ronald Baecker, Stuart Card, Tom Carey, Jean Gasen, Marilyn Mantei, Gary Perlman, Gary Strong, and William Verplank. 1992. *ACM SIGCHI curricula for human-computer interaction*. ACM.

[12] Zayira Jordan, Jose Abdelnour Nocera, Anicia Peters, Susan Dray, and Stephen Kimani. 2016. A Living HCI Curriculum. In *Proceedings of the First African Conference on Human Computer Interaction (AfriCHI'16)*. ACM, New York, NY, USA, 229–232. https://doi.org/10.1145/2998581.2998623

[13] Claudia López, Cleidson R. B. de Souza, Laura S. Gaytán-Lugo, and Francisco J. Gutierrez. 2019. CSCW Research @ Latin America. Retrieved June 9, 2019 from http://interactions.acm.org/blog/view/cscw-research-latin-america

[14] Jaime Muñoz Arteaga, Juan González, and Alfredo Sánchez. 2015. *La Interacción Humano-Computadora en México*. Pearson.

[15] Anicia Peters, Zayira Jordan, Luiz Merkle, Mario Moreno Rocha, Jose Abdelnour Nocera, Gerrit C. van der Veer, Susan Dray, Jennifer Preece, and Elizabeth Churchill. 2016. Teaching HCI: A Living Curriculum?. In *Proceedings of the First African Conference on Human Computer Interaction (AfriCHI'16)*. ACM, New York, NY, USA, 267–270. https://doi.org/10.1145/2998581.2998618

[16] Mariano Pimentel and Hugo Fuks. 2012. *Sistemas Colaborativos (Collaborative Systems, in Portuguese)*. Elsevier.

[17] Pedro C. Santana-Mancilla, Miguel A. Rodriguez-Ortiz, Miguel A. Garcia-Ruiz, Laura S. Gaytan-Lugo, Silvia B. Fajardo-Flores, and Juan Contreras-Castillo. 2019. Teaching HCI Skills in Higher Education through Game Design: A Study of Students' Perceptions. *Informatics* 6, 2 (may 2019), 22. https://doi.org/10.3390/informatics6020022

[18] SIGCHI. 2017. 2011-2014 Education Project. https://sigchi.org/2017/01/ 2011-2014-education-project/

[19] Olivier St-Cyr, Craig M. MacDonald, and Elizabeth F. Churchill. 2019. EduCHI 2019 Symposium: Global Perspectives on HCI Education. In *Extended Abstracts of the 2019 CHI Conference on Human Factors in Computing Systems (CHI EA '19)*. ACM, New York, NY, USA. https://doi.org/10.1145/3290607.3298994

[20] Olivier St-Cyr, Craig M. MacDonald, Elizabeth F. Churchill, Jenny J. Preece, and Anna Bowser. 2018. Developing a Community of Practice to Support Global HCI Education. In *Extended Abstracts of the 2018 CHI Conference on Human Factors in Computing Systems (CHI EA '18)*. ACM, New York, NY, USA. https://doi.org/10.1145/3170427.3170616